\documentclass[10pt,journal,twocolumn,twoside]{IEEEtran}
\usepackage{my_packages}
\usepackage{soul}
\soulregister\ac7
\soulregister\cite7

\pgfplotsset{compat=1.18}
\begin{document}

\title{Design and Performance of Enhanced Spread Spectrum Aloha for Unsourced Multiple Access}

\author{Riccardo~Schiavone,~\IEEEmembership{Graduate Student Member,~IEEE,},
        Gianluigi~Liva,~\IEEEmembership{Senior Member,~IEEE,}
        and~Roberto~Garello,~\IEEEmembership{Senior Member,~IEEE}
        
\thanks{R. Schiavone and R. Garello are with the Department of Electronics and Telecommunications, Politecnico di Torino, 10129 Torino, Italy, e-mail: \{riccardo.schiavone,roberto.garello\}@polito.it}
\thanks{R. Schiavone and G. Liva are with the Institute of Communications and Navigation, German Aerospace Center (DLR), 82234 Wessling, Germany, e-mail: \{riccardo.schiavone,gianluigi.liva\}@dlr.de.}
\thanks{This work was supported in part by the project 6G-RIC of Germany (Program of ``Souver\"an. Digital. Vernetz'', Project No. 16KISK022) and in part by the partnership on ‘‘Telecommunications of the Future’’ of Italy (Program ‘‘RESTART’’, Grant No. PE00000001).}}

\maketitle
\thispagestyle{empty}

\markboth
{submitted to IEEE Communications Letters}
{R. Schiavone, G. Liva, and R. Garello: Design and Performance of E-SSA for Unsourced Multiple Access}

\maketitle

\begin{abstract}
We analyze the performance of enhanced spread spectrum Aloha (E-SSA) in the framework of unsourced multiple access (UMAC). The asynchronous, unframed transmission of E-SSA is modified to enable a direct comparison with framed UMAC schemes and with Polyanskiy's achievability bound. The design of E-SSA is tailored to the UMAC setting, resorting to short polar codes and the use of a timing channel to improve the energy efficiency of the protocol. We assess the impact of the
preamble length and of the spreading factor on the system efficiency.
The resulting scheme exhibits simplicity at the transmitter and linear complexity with respect to the number of active users at the receiver, approaching the UMAC achievability bound in close competition with the best known UMAC schemes.
\end{abstract}

\begin{IEEEkeywords}
Random access, enhanced spread spectrum Aloha, massive machine type communications, polar codes.
\end{IEEEkeywords}

\IEEEpeerreviewmaketitle

\section{Introduction}

\IEEEPARstart{A}{ddressing} the connectivity of a massive number of intermittently transmitting devices---a setting often referred to as \ac{MRA}---poses a significant challenge for the upcoming generations of wireless networks. 
This challenge has garnered increased attention in recent years due to emerging \ac{IOT} applications. The interest in \ac{RA} protocols is shared both in terrestrial \cite{Bockelmann16} and non-terrestrial networks \cite{Cioni18}.

While the design of efficient (coordinated) \ac{MAC} protocols is a well-studied problem in communication systems, the \ac{MRA} setting reveals several crucial aspects that require new solutions  \cite{polyanskiy_perspective_2017}. These include \acl{FBL}  effects \cite{polyanskiy_channel_2010}, the fact that only a small fraction of the users is active at a given moment, and the absence of pre-agreed resource allocation with the base station. From an information theoretic viewpoint, the latter point can be captured by constraining the users to adopt the same codebook, leading to the definition of \ac{UMAC} protocols. The development of an \ac{UMAC} achievability bound \cite{polyanskiy_perspective_2017} provided a fundamental benchmark, triggering the design of new schemes capable of approaching the bound \cite{pradhan2020polarspreading,pradhan2022sparseIDMA,marshakov2019polarIRSA,Truhachev21,Fengler21,Duman24}.

In this paper, we explore the design and the performance of \ac{ESSA} \cite{del_rio_herrero_high_2008} in the \ac{UMAC} framework. \ac{ESSA} is based on spread Aloha \cite{Abr90:SpreadAloha}, and it is a purely asynchronous protocol that combines Aloha \cite{abramson_aloha_nodate} with direct sequence spread spectrum to suppress multiuser interference.  \ac{ESSA} improves the multipacket reception capability of spread Aloha by canceling the interference contribution of decoded packets. The design of \ac{ESSA} is based on the 3GPP W-CDMA air interface, and it is optimized to harvest the gains that \ac{SIC} can provide.
Due to its outstanding performance and lean transmitter/receiver design, \ac{ESSA} emerged as a high-efficiency \ac{MRA} solution, currently adopted in satellite \ac{IOT} networks \cite{etsi_smim}. 

The objective of this letter is two-fold. On one hand, we aim at characterizing the performance of \ac{ESSA} in the \ac{UMAC} setting, showing how a well-established random access protocol that is already deployed in real systems can yield a performance that competes with state-of-the-art \ac{UMAC} solutions. On the other hand, we address specific \ac{ESSA} design choices that stems from the adaptation to the \ac{UMAC} setting, with emphasis on the short packet transmission regime.

To the best of our knowledge no effort has been put to analyze the performance of \ac{ESSA} in the \ac{UMAC} setting. An obstacle to this task is represented by the asynchronous/unframed nature of \ac{ESSA}, which collides with the fixed frame size analysis of \cite{polyanskiy_perspective_2017}. We circumvent this problem by casting a wrap-around version of \ac{ESSA}. 
 By doing so, we can provide a fair basis to compare \ac{ESSA} with finite block length limits \cite{polyanskiy_perspective_2017} and with advanced \ac{UMAC} schemes. The focus is on the \ac{GMAC} channel, which is a realistic model for satellite communications.  We show that a judicious design of \ac{ESSA} allows approaching the bounds up to moderate-size user populations, competing with some of the best known schemes \cite{pradhan2020polarspreading}. 
Considering the transmission of small data units (in the order of a few tens of bits), the result is achieved by modifying the original \ac{ESSA} design according to the following observations. First, the adoption of codes that performs close to finite-length over single-user channels \cite{polyanskiy_channel_2010} allows keeping the gap from the \ac{UMAC} bound \cite{polyanskiy_perspective_2017} small, at low channel loads. Second, the use of protocol information \cite{Gal76:protocol} to carry part of the message content improves the energy efficiency.\footnote{This observation was exploited, for example, in \cite{pradhan2022sparseIDMA,Truhachev21} by mapping a portion of the message onto the preambles / signature sequences envisaged by the scheme, in contrast with the random preamble choice adopted in the 5GNR random access protocol \cite{5GNR16}.}
To address the first point, we employ polar codes \cite{arikan_channelpolarization_2009} concatenated with an outer \ac{CRC}, together with \ac{SCL} decoding \cite{tal2015listdecodingpolar,li2012adaptive} at the receiver side. For the second point, we make use of the \emph{timing channel} associated with the packet transmission time. The use of timing channels to convey information was pioneered in \cite{Gal76:protocol,Ana96:bitsQ} (see also \cite{Eph98:union} for an insightful review of the topic). In the context of random access protocols, timing channels were used in \cite{Gal13:TCAloha} to improve the efficiency of slotted Aloha. In our construction, the timing channel is used to improve the error detection capability of the channel decoder, thus reducing the false alarm rate (which can have a detrimental effect on the \ac{SIC} algorithm performance).
The resulting scheme retains the simple transmitter structure of the original \ac{ESSA}, with a decoding complexity that scales linearly in the number of active users.

\section{Preliminaries}

We denote random variables with capital letters ($X$), and their realizations with lower-case letters ($x$). We define $[n] = \{0,1,\ldots,n-1\}$. The order-$2$ finite field is denoted by $\mathbb{F}_2$.

We consider the transmission of $\ka$ active users over the \ac{GMAC}. The $\ka$ active users transmit simultaneously their message $\bm{X}_i$, with $i=1,...,K_a$, in the same frame of length $n$ real c.u., through a channel affected by \ac{AWGN}. The received vector $\bm{Y}$ is 
\[
\bm{Y} = \bm{X}_1 + \bm{X}_2 + ... + \bm{X}_{K_a} + \bm{Z},\ \bm{Z}\sim\mathcal{N}(\bm{0}_n,\sigma^2 \bm{I}_n)
\]
where $\bm{Z}$ is an $n$-dimensional Gaussian vector with $\bm{0}_n$ (the all-zero vector of length $n$) mean and $\sigma^2 \bm{I}_n$ covariance matrix, and where $\bm{I}_n$ is the identity matrix of size $n\times n$ and $\sigma^2$ is the noise variance. All $\bm{X}_i$ and $\bm{Y}$ have length $n$.  We enforce the power constraint $||\bm{X}_i||_2^2 \leq nP$ for $i = 1, 2, \ldots, \ka$. Following the \ac{UMAC} setting, all messages are select from a common codebook $\code$ with cardinality $|\code|=2^K$, i.e., each message carries $K$ bits of information. The per-user \ac{SNR} is denoted as $E_b/N_0$, where $E_b$ is the energy per information bit, and $N_0$ is the single-sided noise power spectral energy. We have
\[
\frac{E_b}{N_0} = \frac{nP}{2 K \sigma^2}.
\]
Upon observing $\bm{Y}$, the decoder outputs a list $\dec(\bm{Y})$ containing $\ka$ messages from $\code$. The \ac{PUPE} is defined as 
\[
\PUPE := \frac{1}{\ka}\sum_{i=1}^{\ka} \prob{\bm{X}_i \notin \dec(\bm{Y})}.
\]
\subsection{Polar Codes}

An $(N_p,K_p)$ polar code \cite{arikan_channelpolarization_2009}  is specified by the set of indices of the information bits $\mathcal{A}\subseteq [N_p]$.
We denote by $\mathcal{F} = [N_p]\backslash\mathcal{A}$ the set of \textit{frozen} bits. From an information vector $\bm{u} \in \mathbb{F}_2^{K_p}$, the corresponding codeword $\bm{c}$ is obtained by mapping the $K_p$ bits in $\bm{u}$ to the $\mathcal{A}$ positions of a binary vector $\bm{v}$ and freeze to zero the bits of $\bm{v}$ with index in $\mathcal{F}$. The codeword is generated as $\bm{c} = \bm{v}\bm{F}^{\otimes m}$
where $\bm{F}^{\otimes m}$ denotes the $m$-fold Kronecker power of Arikan's $2\times 
2$ polarization kernel.
The blocklength $N_p$ is equal to $2^m$. The performance of polar codes under \ac{SCL} decoding can be improved by concatenating an inner polar code with an outer high-rate code. Following \cite{tal2015listdecodingpolar}, we employ an outer $(K_p,K)$ \ac{CRC} code, resulting in a concatenated \ac{CRC}-polar code with blocklength $N_p$ and dimension $K$. In the rest of the paper, we adopt the polar code design of the 5GNR standard \cite{tal2015listdecodingpolar}, with the \ac{CRC} code polynomial of degree $11$. We refer to this code construction as CRC-aided polar (CA-polar) code. Following the standard, we perform rate matching (puncturing) to obtain a $(N,K)$ CA-polar code of prescribed blocklength $N\leq N_p$. We use adaptive \ac{SCL} \cite{li2012adaptive} decoding with maximum list size $256$.

\section{Protocol Design}

In this section, we describe the design of the scheme. We first define the transmission protocol (Section \ref{sec:tx}), including the modifications applied to \ac{ESSA} to comply with the fixed frame size setting of \cite{polyanskiy_perspective_2017}. We then outline the algorithms performed at the receiver (Section \ref{sec:rx}). Finally, we discuss the role played by the timing channel in Section \ref{sec:timing}.

\subsection{Message Transmission}\label{sec:tx}

The transmission of a message resembles closely the one employed originally by \ac{ESSA} \cite{del_rio_herrero_high_2008}, with a few important differences.  First, we adopt a wrap-around approach to turn the continuous, unframed transmission behavior of \ac{ESSA} into a framed one. This choice is dictated by the interest in comparing the protocol performance with available performance bounds, which rely on a fixed frame size \cite{polyanskiy_perspective_2017}. Second, owing to their excellent performance in the short blocklength regime, we employ the CA-polar codes in place of the turbo codes used in \cite{del_rio_herrero_high_2008}. Furthermore, we define the transmission time of a packet within a frame to be a function of the message. This information will be used, at the receiver end, to improve the error detection capability of the CA-polar code.

Transmission takes place by encoding the information message with the $(N,K)$ CA-polar code, by spreading the codeword through a long spreading sequence $\bm{b}$ with a period larger than the spreading factor $s$, and by appending to the spread packet a preamble $\bm{p}$. The transmission proceeds with a start time obtained by hashing the information message. As for the original \ac{ESSA} design, both the preamble and the spreading sequence are unique, i.e., all users employ the same preamble and the same spreading sequence. Let us denote by $\bm{p} = (p_0, p_1, \ldots, p_{L_0-1})$ the length-$L_0$ preamble, with symbols $p_i \in \{-1,+1\}$ for $i=0,\ldots,L_0-1$. Furthermore, we denote by $\bm{b} = (b_0, b_1, \ldots, b_{L-1})$ the spreading sequence with spreading factor $s = L/N$ (the length of the spreading sequence is a multiple of the CA-polar code blocklength $N$). As for the preamble, the spreading sequence is binary ($b_i \in \{-1,+1\}$ for $i=0,\ldots,L-1$). 
We partition the spreading sequence into $N$ disjoint blocks of $s$ symbols each as $\bm{b} = (\bm{b}_0, \bm{b}_1, \ldots, \bm{b}_{N-1})$. Given the spreading sequence $\bm{b}$, for $\bm{c} \in \mathbb{F}_2^N$ we introduce the spreading function
\begin{equation*}
    \spread(\bm{c};\bm{b}) := \left((-1)^{c_0}\bm{b}_0, (-1)^{c_1}\bm{b}_1, \ldots,(-1)^{c_{N-1}}\bm{b}_{N-1}  \right).
\end{equation*}
We finally denote by $\hash: \mathbb{F}_2^K \mapsto [n]$ a uniform hash function.
The transmission of a message proceeds as follows.

\medskip

\begin{itemize}
\item[1.] \emph{Polar Encoding:} The $K$-bits information vector $\bm{u}$ is encoded via the $(N,K)$ CA-polar encoder, resulting in the $N$-bits codeword $\bm{c}$.
\item[2.] \emph{Spreading:} The $N$-bits codeword is spread via the sequence $\bm{b}$, resulting in the length-$L$ vector
$\bm{d} = \spread(\bm{c};\bm{b})$.
\item[3.] \emph{Preamble Insertion and Zero-Padding:} The preamble $\bm{p}$ is appended to the vector $\bm{d}$. The resulting length-$(L_0+L)$ vector is padded with $(n-(L_0 + L))$ zeros, resulting into length-$n$ vector
$\bm{x}' = (\bm{p},\bm{d},0, 0 , \ldots, 0)$.
\item[4.] \emph{Hashing and Time Selection:} The transmission time is obtained by hashing the information message as $t = \hash(\bm{u})$.
\item[5.] \emph{Delay Shift and Transmission}. The transmitted vector $\bm{x}$ is given by the circular-right shift of $\bm{x}'$ by $t$ positions.
\end{itemize}

\medskip

\begin{figure}
    \centering
    \includegraphics[width=0.95\columnwidth]{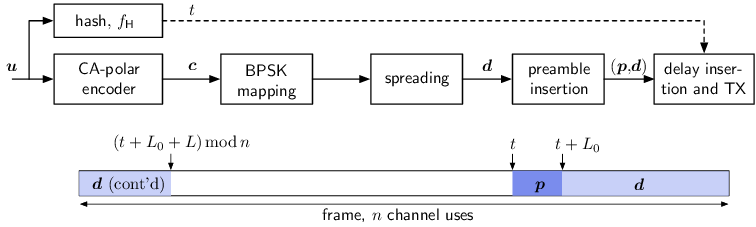}
    \vspace*{-0.3cm}
    \caption{Structure of the transmission chain.}
    \vspace*{-0.2cm}
    \label{fig:transmitter}
\end{figure}

A sketch of the transmitter is shown in Figure \ref{fig:transmitter}. Due to the choice of the binary alphabet, the average power is $P = (L_0+L)/n$. Note that, the system employs a pre-defined preamble and a single, pre-defined spreading sequence, i.e.,  each user employs the same preamble and the same spreading sequence, which are known at the receiver. Remarkably, due to the asynchronous nature of spread Aloha, the use of a single spreading sequence is sufficient to reduce multiuser interference: users that transmit at different times will experience a mutual interference that closely resembles the one of a synchronous spread spectrum system with random spreading sequences assigned to the different users \cite{abramson_aloha_nodate}.

\subsection{Detection and Decoding}\label{sec:rx}
The receiver runs iteratively. In each iteration, a list of candidate start-of-packet positions is produced through a preamble search. For each candidate position, a decoding attempt is performed via adaptive \ac{SCL} decoding \cite{li2012adaptive}. If the decoder list contains a codeword that satisfies the \ac{CRC} code constraints, the associated information message is hashed with the function $\hash$, and the output is checked with the start time of the detected preamble. If the output of the hash function matches the start time, the decision is considered correct and the decoded packet interference contribution is removed from the received signal. The behavior of the receiver is detailed next.

\medskip

\noindent \textbf{Initialization:} The iteration count is set to $I=0$.

\smallskip

\noindent \textbf{Iterative Decoding and \ac{SIC}:} At the $\ell$th iteration, the following steps are performed: 
\begin{itemize}
    \item[1.] \emph{Preamble Detection:} The correlation between the preamble and the  vector $\bm{y}$ is computed at each $t \in [n]$ as 
    \[
        \Lambda_t = \sum_{j=0}^{L_0-1} p_j y_{(j+t)\,\mathrm{mod}\, n}.
    \]
    The times associated to the $W$ largest values of $\Lambda_t$ are stored in the list $\mathcal{T}$.  
    \item[2.] \emph{Despreading, Decoding, and SIC:} For each $t \in \mathcal{T}$
    \begin{itemize}
        \item[a.] The observation vector $\bm{r}$ is extracted as 
        $\bm{r}=\left(y_{t'},\ldots,y_{t''}\right)$
        where $t' = (t+L_0)\, \mathrm{mod}\, n$ and $t'' = (t+L_0+L-1)\, \mathrm{mod}\, n$. 
        \item[b.] A soft-estimate of the $j$th codeword bit is obtained by matched filtering (despreading) as 
        \[\tilde{r}_j = \frac{1}{s}\sum_{k=js}^{(j+1)s-1}  r_k b_{k}.\]
        \item[c.] The soft-estimate vector $\tilde{\bm{r}} = (\tilde{r}_0, \tilde{r}_1, \ldots, \tilde{r}_{N-1})$ is input to the polar code adaptive \ac{SCL} decoder \cite{li2012adaptive}. The decoder returns a valid codeword $\hat{\bm{c}}$ or an erasure flag (if no polar code word in the \ac{SCL} decoder list satisfies the \ac{CRC} code constraints).
        \item[d.] If the decoder outputs a decision $\hat{\bm{c}}$, the corresponding information message $\hat{\bm{u}}$ is hashed generating the time index $\hat{t} = \hash(\hat{\bm{u}})$. If $\hat{t} = t$, then the message $\hat{\bm{u}}$ is deemed as correct and it is included in the output list $\dec(\bm{y})$. 
        \item[e.] If the check at point 2.d succeeds, the message is re-encoded according to the transmission procedure described in Section \ref{sec:tx}, resulting in the vector $\bm{x}(\hat{\bm{u}})$, which is subtracted from the vector $\bm{y}$
        \[
        \bm{y} \leftarrow \bm{y} - a \bm{x}(\hat{\bm{u}})
        \]
        where $a$ is the estimate of the channel amplitude\footnote{Note that the detection/decoding algorithm described in this section does not make use of any prior information on the unitary channel amplitude.} obtained by normalizing the soft-correlation between $\bm{y}$ and $\bm{x}(\hat{\bm{u}})$  by the vectors' length ($L_0 + L$).
    \end{itemize} 
\end{itemize}

Steps $1$ and $2$ are iterated for a maximum number of iterations $I_\mathrm{max}$. As early stopping criterion, the process ends if---within an iteration---no decoding attempt succeeds according to the test described at step 2.d.
It is important to remark that, if we fix $I_\mathrm{max}$ and we scale $W$ with the number of active users $\ka$, the detection/decoding algorithm outlined above entails a complexity that is linear in $\ka$.

\subsection{Error Detection via Timing Channel}\label{sec:timing}
The use of the timing channel for error detection, as described in the algorithm described in Section \ref{sec:rx} (step 2.d) allows one to drastically reduce the \emph{false alarm rate}, i.e., the rate at which the decoder outputs packets that were not transmitted. False alarms have a detrimental effect on  the efficiency of the system, since they introduce artificial interference through the \ac{SIC} process. It is hence of paramount importance to keep the  false alarm rate some orders of magnitude lower than the target \ac{PUPE}. In our simulations, for instance, with $125$ active users, we counted exactly three false alarms in  $800$ simulated frames, whereas with $K_a=100$ and $K_a=75$ no false alarms were detected. As an alternative to the use of the timing channel, one may improve the error detection capability of the \ac{SCL} decoder. The result can be achieved by limiting the \ac{SCL} list size, or by introducing a stronger \ac{CRC}. In either case, the price to be paid is an deterioration of the error correction capability of the decoder, resulting in a loss of coding gain, thus in a loss of energy efficiency. An obvious question relates to the practicality of using the timing information in a real system. Answering thoroughly this question goes beyond the scope of this paper. However, a possible direction to explore is the use of a beacon signal (transmitted periodically by the base station) to announce the start of frame, which is used by the terminals as reference to compute the transmission time. Uncertainties related to the delay incurred by the transmission may be taken into account by relaxing the criterion of step 2.d in the algorithm of Section \ref{sec:rx}: a decoding attempt may be deemed successful if $\hat{t} \in [t-\delta,t+\delta]$, where $\delta$ is a suitably chosen maximum delay offset. For example, in our numerical test we did not experience any visible degradation in error detection capability with values of $\delta$ up to $150$ channel uses. Closed-loop timing control mechanisms, as the ones implemented by the 5GNR \cite{5GNR16} and by the DVB-RCS standards \cite{ETSI09b}, may be also considered as means to keep the timing error within a pre-defined range.

\section{Numerical Results}\label{sec:results}
In this section, we provide numerical results obtained via Monte Carlo simulations.
The results are obtained for a frame size $n=30000$. A $(1000,100)$ CA-polar code is used, and the number of iterations is set to $I_{\mathrm{max}}=50$. The target \ac{PUPE} is $\epsilon^{\star} = 5 \times 10^{-2}$.
For the simulations, we used a single randomly-generated spreading sequence, with symbols that are independent and uniformly distributed in $\{\pm 1\}$.

\begin{figure}
    \centering
    \includegraphics[width=\columnwidth]{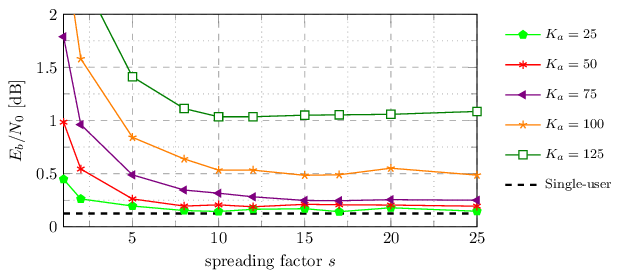}
    \vspace*{-0.7cm}
    \caption{Effect of the spreading factor $s$ on the minimum $E_b/N_0$ required for $\epsilon^{\star}=5\times 10^{-2}$. Genie-aided preamble detection.}\vspace*{-0.2cm}
    \label{fig:EbN0_vs_spreading_factor}
\end{figure}

\begin{figure}
    \centering
    \includegraphics[width=\columnwidth]{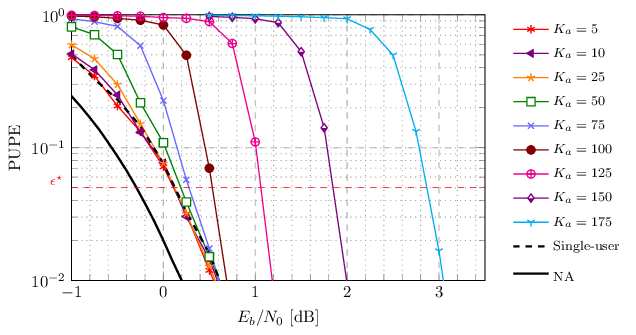}
    \vspace*{-0.7cm}
    \caption{Effect of the channel load $K_a$ on the \ac{PUPE} in a system with no preamble. Spreading factor $s=25$. Genie-aided preamble detection.}\vspace*{-0.3cm}
    \label{fig:threshold_effect}
\end{figure}

A first set of results deals with the choice of the spreading factor. In Figure \ref{fig:EbN0_vs_spreading_factor}, we report the \ac{SNR} required to achieve the target \ac{PUPE} as a function of the spreading factor. The results are provided for various numbers of active users, and are obtained under genie-aided preamble detection. That is, the preamble is omitted. The chart shows how small spreading factors tend to penalize the performance. For larger spreading factors, and at low channel loads (small $\ka$), the required \ac{SNR} closely approaches the one of the single-user case. The result holds up to $K_a = 75$. Above this value, there is a rapid deterioration of the performance. This behavior is analyzed in Figure \ref{fig:threshold_effect}. Here, the \ac{PUPE} is depicted as a function of the \ac{SNR}, for various numbers of active users. The spreading factor is set to $25$. On the same chart, the normal approximation \cite{polyanskiy_channel_2010} for $(1000,100)$ codes is provided, as well as the single-user $(1000,100)$ CA-polar code performance. The \ac{PUPE} approaches the single-user performance at low enough error probability. The \ac{SNR} at which the convergence to the single user case varies with the number of active users. When the number of active users is sufficiently low, the \ac{PUPE} tightly matches the single-user probability of error already at error probabilities larger than the target $\epsilon^\star$. As the number of users grows, the convergence happens at \ac{PUPE} values that are below $\epsilon^\star$, giving rise to a visible increase of the required \ac{SNR}. An analysis of this phenomenon, which is common in multiuser systems, was provided in \cite{kowshik2021fundamental}.


\begin{figure}
    \centering
    \includegraphics[width=\columnwidth]{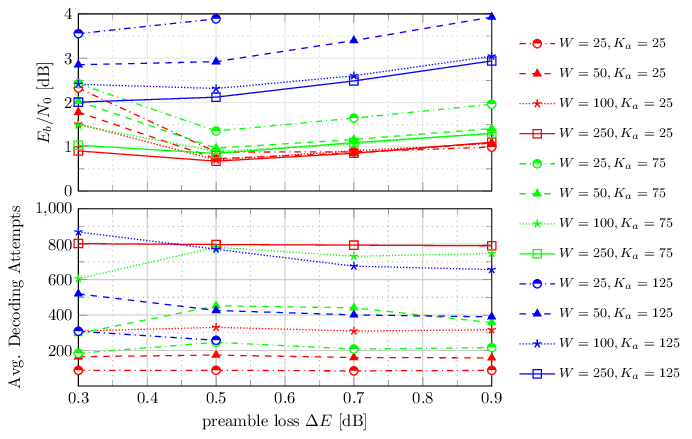}
    \vspace*{-0.6cm}
    \caption{Minimum $E_b/N_0$ (above) and average number of decoding attempts (below) for $\epsilon^{\star}=5\times 10^{-2}$ as a function of the preamble loss, $s=15$.}
    \vspace*{-0.20cm}
    \label{fig:complexity_EbN0}
\end{figure}

A second set of results analyzes the impact on the system performance of the number $W$ of start-of-packet candidates used at step 1 of the algorithm described in Section \ref{sec:rx}. The results, obtained for $s=15$ and for various preamble lengths $L_0$, are depicted in Figure \ref{fig:complexity_EbN0}  in terms of \ac{SNR} required to achieve the target \ac{PUPE} as a function of the energy overhead introduced by the preamble, given by $\Delta E := 1 + L_0/L$. As one would expect, a larger preamble overhead (larger $L_0$) enables an accurate preamble detection already with small values of $W$, resulting also in a smaller average number of decoding attempts. However, the result comes at the expense of an increase in the energy cost entailed by the preamble. Operating the system with lower preamble lengths allows to limit the energy loss. Anyhow, the preamble miss-detection probability can be kept low only by increasing $W$, with an obvious implication in complexity due to the larger number of decoding attempts. The analysis highlights the trade-off that exists at fixed \ac{SNR}/\ac{PUPE} between a reduction of the preamble overhead and the corresponding increase of decoding complexity, suggesting that the preamble length shall be selected by finding an acceptable compromise between energy overhead and size of the list of start-of-packet candidates.

\begin{figure}
    \centering
    \includegraphics[width=\columnwidth]{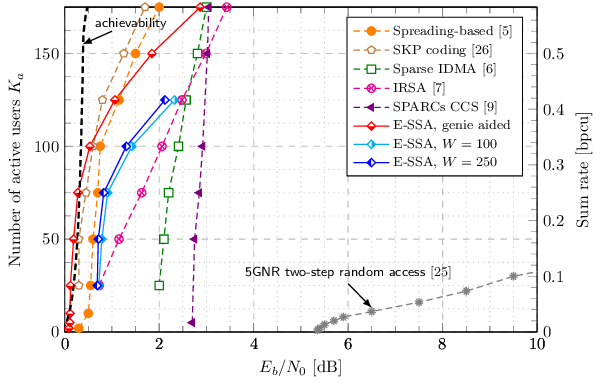}
    \vspace*{-0.65cm}
    \caption{$E_b/N_0$ required to achieve the target $\PUPE^\star=5\times 10^{-2}$.}
    \vspace*{-0.30cm}
    \label{fig:comparison_other_schemes}
\end{figure}

Figure \ref{fig:comparison_other_schemes} provides a comparison of \ac{ESSA} with the UMAC achievability bound and several \ac{UMAC} schemes.
The results are obtained by setting the \ac{ESSA} preamble length to $L_0 = 3050$ and a spreading factor $s=25$, yielding an energy overhead $\Delta E \approx 0.5$ dB. The \ac{ESSA} performance is provided for $W=100$ and for $W=250$. The performance with genie-aided preamble detection is given as reference, together with the \ac{UMAC} achievability bound from \cite{polyanskiy_perspective_2017}. The comparison includes the 5GNR two-step random access procedure \cite{5GNR16} under \ac{SIC} with parameters that have been chosen to match the simulation setup \cite{liva2024UMAC}, the sparse Kronecker-product coding of \cite{han2021sparse}, the synchronous spread spectrum scheme of \cite{pradhan2020polarspreading}, the enhanced irregular repetition slotted Aloha scheme of \cite{marshakov2019polarIRSA}, sparse interleave-division multiple access \cite{pradhan2022sparseIDMA} (which exploits joint multiuser decoding), and sparse regression codes / coded compressive sensing \cite{Fengler21}. The performance of \ac{ESSA} shows to be competitive, especially in the moderate load regime (up to $100$ active users), requiring only single-user detection and decoding blocks at the receiver side. Among the competitors, only the schemes \cite{pradhan2020polarspreading,han2021sparse} outperform \ac{ESSA} over the entire channel load range, with a gain that is nevertheless limited to $0.2$ dB for \cite{pradhan2020polarspreading} and $0.5$ dB for \cite{han2021sparse} up to $75$ users.

\section{Conclusion}
We analyzed the performance of enhanced spread spectrum Aloha (E-SSA) in the framework of unsourced multiple access (UMAC). The asynchronous, unframed transmission of E-SSA has been modified to enable a comparison with framed UMAC schemes. We have improved the energy efficiency of the scheme by introducing a short polar code and a timing channel. We have shown how the design of the different components of E-SSA affects the receiver complexity and the energy efficiency of the system. Results show that a careful design of E-SSA yields a performance that is competitive with state-of-the-art UMAC schemes, with a simple transmitter and a linear receiver complexity in the number of active users.

\section*{Acknowledgments}

The authors would like to thank Monica Visintin and they are also grateful to Yury Polyanskiy for providing the software used to compute the achievability bound of \cite{polyanskiy_perspective_2017}.

\bibliography{IEEEabrv,my_library}
\bibliographystyle{IEEEtran}

\end{document}